\documentclass[aps,amsmath,twocolumn,amssymb,titlepage,superscriptaddress,10pt]{revtex4-1}
\usepackage[T1]{fontenc}
\usepackage[utf8]{inputenc}
\usepackage{amsmath}
\usepackage{braket}
\usepackage{amsfonts}
\usepackage{comment}
\usepackage{graphicx}
\usepackage[breaklinks,colorlinks,bookmarks=false,citecolor=blue,linkcolor=red,urlcolor=blue]{hyperref}
\usepackage[capitalize]{cleveref}
\usepackage{amssymb}
\usepackage{amsmath}
\usepackage{mathtools}
\begin{document}

\title{Thermal critical points from competing singlet formations\\ in fully frustrated bilayer antiferromagnets}

\author{Lukas Weber}
\email{lweber@flatironinstitute.org}
\affiliation{Center for Computational Quantum Physics, Flatiron Institute, 162 5th Avenue, New York, NY 10010}
\affiliation{Max Planck Institute for the Structure and Dynamics of Matter, Luruper Chaussee 149, 22761 Hamburg, Germany}
\author{Antoine Yves Dimitri Fache}
\affiliation{Institute of Physics, Ecole Polytechnique Fédérale de Lausanne (EPFL), CH-1015 Lausanne,
Switzerland}%
\author{Frédéric Mila}
\affiliation{Institute of Physics, Ecole Polytechnique Fédérale de Lausanne (EPFL), CH-1015 Lausanne,
Switzerland}%
\author{Stefan Wessel}
\affiliation{%
Institute for Theoretical Solid State Physics, RWTH Aachen University,
JARA Fundamentals of Future Information Technology, and
JARA Center for Simulation and Data Science, 52056 Aachen, Germany
}%

\begin{abstract}
We examine the ground-state phase diagram and  thermal phase transitions in a plaquettized fully frustrated bilayer spin-1/2 Heisenberg model. Based on a combined analysis from sign-problem free quantum Monte Carlo simulations, perturbation theory and free-energy arguments, we identify a first-order quantum phase transition line that separates two competing quantum-disordered
ground states with dominant singlet formations on inter-layer dimers and plaquettes, respectively. At finite temperatures, this line extends to form  a wall of first-order thermal transitions, which terminates in  a line of thermal critical points. From  a perturbative approach in terms of an effective Ising model description,  we identify a quadratic suppression of the critical temperature scale in the strongly 
plaquettized region. Based on free-energy arguments we furthermore obtain the full  phase boundary of the low-temperature dimer-singlet regime, which agrees well with the quantum Monte Carlo data. 

\end{abstract}

\maketitle

\def\JD{J_\text{D}}
\def\JP{J_\text{P}}
\def\nD{n_\text{D}}

\newcommand\ketbra[2]{\ket{#1}\!\bra{#2}}

\section{Introduction}
Geometric frustration in  quantum magnets   can give rise to a variety of non-classical ground states, including quantum-disordered states that are  dominated  by the formation of local spin singlets on particular sub-clusters~\cite{Richter2004,Balents2010,HFMbook,DIEPbook}. Examples include dimer singlet and plaquette singlet states, where spin singlets form predominantly among two- and four- spin sub-clusters, respectively.  
Such quantum-disordered regions are often separated by discontinuous (first-order) quantum phase transition lines in the parameter space of the system. Thermal fluctuations may replace the discontinuous quantum phase transition by a continuous thermal crossover between these different regimes, but it is also possible that the discontinuous behavior remains stable at low temperatures. 
In recent years, several instances were indeed reported in strongly frustrated quantum magnets in which a  discontinuous quantum phase transition line extends beyond the zero-temperature limit, forming a boundary of first-order thermal transitions in the thermal phase diagram~\cite{Stapmanns2018,Jimenez2021,Weber2021}. It was found that such a “wall of discontinuities”  terminates along a line of thermal critical points. In the two-dimensional (2D) models studied in these references, these critical points belong to the universality class of the 2D Ising model. This reflects the fact that a single scalar quantity is sufficient to distinguish the phases, hence to describe the critical fluctuations at the thermal critical points~\cite{Stapmanns2018}.

A prominent example for this  scenario is provided by the layered compound SrCu${}_2$(BO${}_3$)${}_2$, a material that received increasing attention recently in the field of frustrated quantum magnetism~\cite{Jimenez2021}:
In SrCu${}_2$(BO${}_3$)${}_2$, a pressure-induced discontinuous quantum phase transition takes place between a  dimer singlet product phase and a plaquette singlet quantum-disordered phase at about 20 kbar. The low-temperature first-order transition line was found to terminate at a critical point at a temperature of about 4K, i.e., well below the scale of the magnetic exchange interactions in this system. Upon approaching the critical point, the specific heat furthermore exhibits characteristic critical enhancement, as in the 2D Ising model. 

Prior to its experimental observation in SrCu${}_2$(BO${}_3$)${}_2$, this  physics was identified~\cite{Stapmanns2018} in a related basic 2D model of strongly frustrated quantum magnetism, the fully-frustrated bilayer (FFB) spin-1/2 Heisenberg antiferromagnet (AFM)~\cite{MullerHartmann2000}. In the FFB, a discontinuous quantum phase transition  takes place between a dimer singlet phase and an AFM ordered phase. Building on recent progress in  designing minus sign-problem free quantum Monte Carlo (QMC) approaches for frustrated quantum magnets~\cite{Honecker2016, Alet2016}, it is now possible to study this quantum phase transition and the critical point that terminates the extended first-order transition line by unbiased and large-scale QMC simulations.

In contrast to the case of SrCu${}_2$(BO${}_3$)${}_2$, the temperature scale of the critical point in the FFB model turns out to be of similar  magnitude as the magnetic exchange interaction strengths. Another difference between the FFB model and SrCu${}_2$(BO${}_3$)${}_2$ is the fact that in the FFB the discontinuous quantum phase transition takes place between an AFM ground state and a quantum-disordered phase, while in SrCu${}_2$(BO${}_3$)${}_2$, the phases on both sides of the quantum phase transition point are non-magnetic and quantum disordered. It would thus be interesting to come up with an example of a discontinuous quantum phase transition between two quantum-disordered regions in a quantum spin model that is accessible to sign-problem free QMC simulations. 

\begin{figure}[t]
\includegraphics{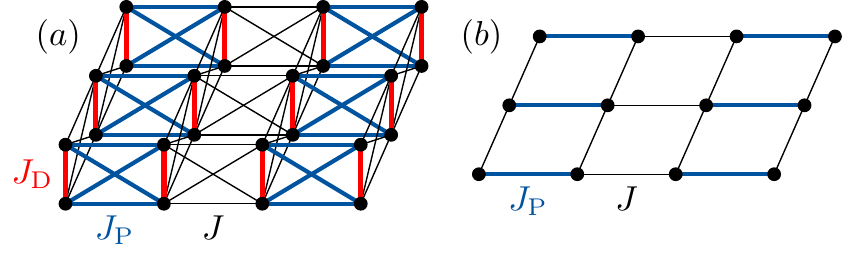}
\caption{(a) The pFFB lattice with interlayer dimer bonds $J_D$ (thick, red), plaquette bonds $J_P$ (thick, blue) and interplaquette bonds $J$ (thin, black). (b) The effective spin-1 model for the pFFB within the dimer spin triplet regime.}
\label{fig:lattice}
\end{figure}

Here, we consider an extension of the original FFB model that exhibits a line of discontinuous quantum phase transitions between two quantum disordered phases with different singlet patterns, and which can be studied by sign-problem free QMC simulations. More specifically, we consider the  plaquettized fully-frustrated bilayer (pFFB) spin-1/2 Heisenberg model, cf. Fig.~\ref{fig:lattice}(a), and defined in detail below. From sign-problem free QMC simulations combined with analytical results from perturbation theory as well as free-energy considerations, we find that in this system the line of discontinuous quantum phase transitions yields a finite-temperature  wall of discontinuities that terminates along a line of 2D Ising critical  points, with a critical temperature that is strongly suppressed with respect to the magnetic exchange couplings, i.e., similar to the case of  SrCu${}_2$(BO${}_3$)${}_2$.

The remainder of this paper is organized as follows: in Sec.~\ref{Sec:Model}, we introduce the pFFB model and present results from sign-problem free QMC simulation of this model in Sec.~\ref{Sec:QMC}. Next, we report our analytical findings in Sec.~\ref{Sec:Analytics}, before giving final conclusions in Sec.~\ref{Sec:Conclusions}.

\section{Model}\label{Sec:Model}

The pFFB model that we consider in the following is a spin-1/2 Heisenberg AFM on the plaquettized bilayer square lattice, shown in Fig.~\ref{fig:lattice}(a). It is defined by the Hamiltonian
\begin{equation}
	H = \JD \sum_{\braket{i,j}_\text{D}} \mathbf{S}_i \cdot \mathbf{S}_j + \JP\!\!  \sum_{\braket{i,j}_\text{P}} \mathbf{S}_i \cdot \mathbf{S}_j + J\!\!\!  \sum_{\braket{i,j}_{{\text{P}-\text{P}}}}\!\!\!\! \mathbf{S}_i \cdot \mathbf{S}_j,
\end{equation}
where $\mathbf{S}_i$ denotes a spin-1/2 degree of freedom on the $i$-th lattice site, and the summations extend (from left to right) over the interlayer dimer bonds, the plaquette bonds and the interplaquette bonds, respectively, cf. Fig.~\ref{fig:lattice}(a). The  four-site unit cell, containing two $\JD$-dimer bonds and four $\JP$ interdimer bonds, is also referred to as  a plaquette in the following. 

The above Hamiltonian can be rewritten in terms of total dimer spin variables. Namely, for each $\JD$-dimer $d$, we define the total dimer spin operator $\mathbf{T}_d=\mathbf{S}_{d,1}+\mathbf{S}_{d,2}$, i.e., the sum of the spin operators of the two sites that belong to the $d$-th dimer. In terms of these operators, the Hamiltonian $H$ reads
\begin{equation}\label{eq:Hdimer}
	H = \JD \sum_{d}\left[ \mathbf{T}_d^2-\frac{3}{4}\right]+ \JP\!\! \sum_{\langle d,d' \rangle_\text{P}} \!\!\mathbf{T}_d\cdot \mathbf{T}_{d'} 
	+\: J\!\!\!\!\! \sum_{\langle d,d' \rangle_{{\text{P}-\text{P}}}} \!\! \!\!\!\!\mathbf{T}_d\cdot \mathbf{T}_{d'},
\end{equation}
where the summations extend (from left to right)  over the interlayer dimers, neighboring dimers coupled by plaquette bonds, and neighboring dimers coupled by interplaquette bonds, respectively. 

This expression makes it clear that $H$ has extensively many   local conserved quantities, namely each total dimer 
spin $\mathbf{T}_d^2$, which we may encode in additional quantum
numbers $T_d$, which take on the values $0$ and $1$ for dimer singlet and triplet states, respectively. In the dimer triplet sector, where $T_d=1$ on all dimers, the Hamiltonian $H$ then describes a spin-1 Heisenberg model on a square lattice with a columnar dimerization pattern, cf. Fig.~\ref{fig:lattice}(b), i.e. the spin-1 columnar dimer square lattice Heisenberg model.  

In several limiting cases, the physics of the pFFB model is readily accessible. If the couplings $\JD$ ($\JP$) dominate, the model will host a dimer (plaquette) singlet  quantum-disordered ground state, denoted DS (PS), in which singlets predominantly form on the $\JD$ dimers ($\JP$ plaquettes), giving rise to a finite triplet excitation gap in both cases. If the coupling $J$ dominates, the system decouples into a system of weakly coupled one-dimensional spin tubes formed by the $J$-bonds (cf. Fig.~\ref{fig:lattice}(a)).
Along $\JP=J$, the pFFB reduces to the original FFB, where, if $J/\JD > 0.42957(2)$~\cite{MullerHartmann2000}, the ground state hosts long-range AFM order, with  each dimer forming an effective $S=1$ degree of freedom, while for lower values of $J$, the FFB resides in the DS phase. In the following, we will examine  the  full phase diagram  of the pFFB model in the antiferromagnetic regime, i.e. assuming all exchange couplings to be positive.

\section{Quantum Monte Carlo results}\label{Sec:QMC}

Even though the Hamiltonian $H$ is strongly frustrated, we can obtain unbiased numerical results for its properties by employing sign-problem free stochastic series expansion (SSE) QMC simulations~\cite{Sandvik1991,Sandvik1999,Syljuasen2002,Alet2005} in the dimer spin basis~\cite{Honecker2016,Alet2016}. Here, we consider systems with periodic boundary conditions, consisting of $L\times L$ unit cells with $N=4 L^2$ spins.

Two basic observables that allow us to distinguish the different phases of the pFFB model are directly accessible in the dimer spin basis: (i) The dimer triplet density $n_\text{D}=\langle N_\text{D}/N\rangle$, where the operator $N_\text{D}$ counts the number of $\JD$-dimers that are in a triplet state, and (ii) the AFM spin structure factor
\begin{equation}
    S(\pi,\pi)=\frac{1}{2 L^2}\sum_{i,j=1}\!\!\: \epsilon_i \epsilon_j \: \langle \mathbf{S}_i\cdot \mathbf{S}_j \rangle,
\end{equation}
where $\epsilon_i=(-1)^{x_i+y_i}$, in terms of the coordinates of lattice site $i$. This observable is susceptible to long-ranged AFM correlations within each of the planes of the bilayer lattice. 

\begin{figure}[t]
	\includegraphics{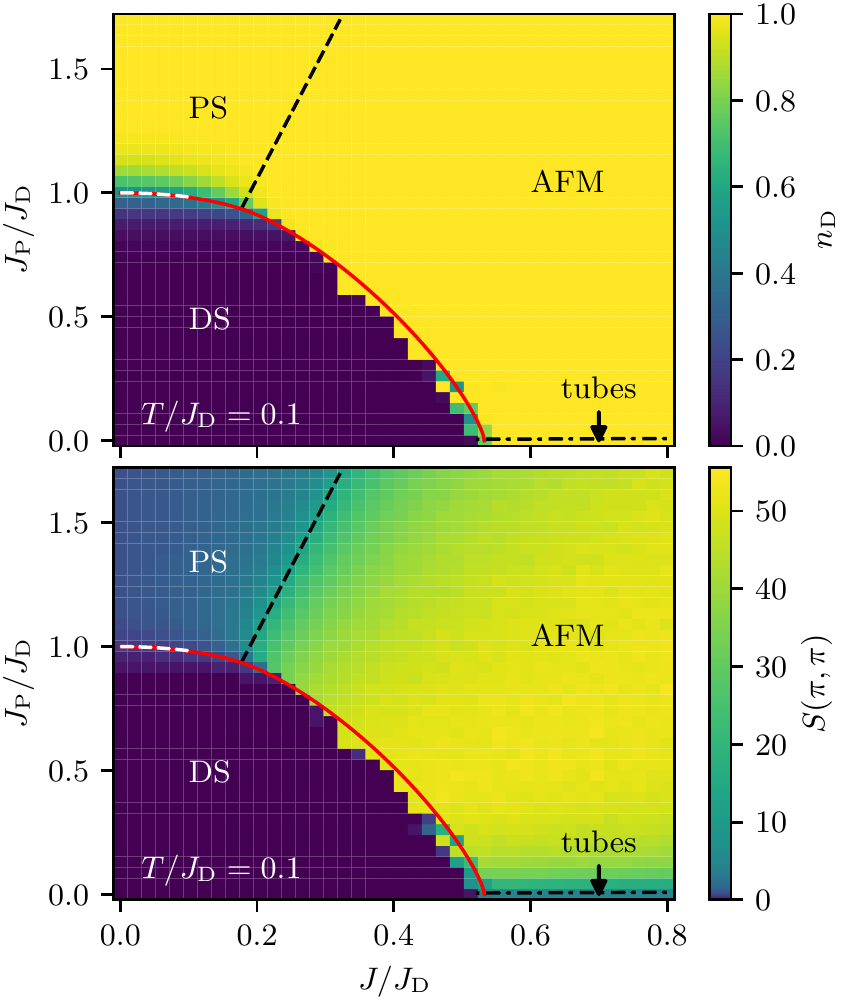}
	\caption{Dimer triplet density $\nD$ (top panel) and  AFM structure factor $S(\pi,\pi)$ (bottom panel) of the pFFB model as a function of $J/\JD$ and $\JP/\JD$ at a fixed temperature of $T/\JD=0.1$ as obtained from QMC for $L=12$. Black lines denote the  boundaries of the AFM phase as obtained from the effective spin-1 model description. Red (white) lines denotes the first-order quantum phase transition line obtained from the free-energy comparison (perturbation theory in the regime $J\ll \JP\approx \JD$). The different regions are labeled by the corresponding ground state phases.}
	\label{fig:ntrip}
\end{figure}

QMC results for both observables are presented at a low temperature of $T/\JD=0.1$ in Fig.~\ref{fig:ntrip} for an $L=12$ system. Both quantities are shown in the parameter plane spanned by the coupling ratios $J/\JD$ and $\JP/\JD$. Combining the data from the two panels, we can readily identify four regimes, denoted DS, PS, AFM, and tubes, which we already introduced above and which all appear as extended regions in the ground-state phase diagram. For low values of both $J/\JD$ and $\JP/\JD$, the dominant $\JD$ coupling forces the system into the DS phase, with very small values of both $n_\text{D}$ and $S(\pi,\pi)$. Outside the DS region, the triplet density $n_\text{D}$ is essentially saturated, and the structure factor $S(\pi,\pi)$ allows us to separate the AFM regime, with a sizeable value of $S(\pi,\pi)$, from both the  PS region (for dominant $\JP$) and the tube phase (for dominant $J$). Since long-range AFM order is restricted to zero temperature, the structure-factor data in Fig.~\ref{fig:ntrip}, taken at low but finite temperature, varies continuously across the corresponding phase transition regimes.  

As seen from the formulation of the Hamiltonian $H$ in terms of the dimer spin operators, cf. Eq.~\eqref{eq:Hdimer}, inside the dimer triplet dominated regime the pFFB model becomes an effective spin-1 Heisenberg model on the columnar dimer square lattice, cf. Fig.~\ref{fig:lattice}(b). The ground state phase diagram of this spin-1 model has been determined by QMC simulations in Ref.~\cite{Matsumoto2001}. From those results we can extract the critical coupling ratios $(J/\JP) = 0.189 20(2)$ and $(J/\JP) \approx 1/0.011\approx 91$ for the continuous quantum phase transitions between the AFM regime and the large-$\JD$ PS and the large-$J$ tube phase, respectively. The black lines in Fig.~\ref{fig:ntrip} indicate these transition lines, which match very well the QMC results. 

Along the line $\JP=J$, where the pFFB  reduces to the original FFB, both quantities exhibit a pronounced jump as the coupling $J$ is tuned across the previously determined position of the DS-to-AFM quantum phase transition at $J/\JD=0.42957(2)$~\cite{MullerHartmann2000}. Indeed, in this regime, the simulation temperature  used for Fig.~\ref{fig:ntrip}  is well below  the critical temperature $T_c\approx 0.22 \JD$~\cite{Stapmanns2018} of the FFB, i.e., at $T=0.1\JD$ the system is driven across the first-order thermal transition line upon increasing $J$, which leads to the sudden jump in both quantities, observed already in Ref.~\cite{Stapmanns2018} (QMC data taken at temperatures beyond $T_c$ instead show a smooth crossover behavior, cf. App.~\ref{app:higherT}). As seen from Fig.~\ref{fig:ntrip}, the sudden change in both quantities remains similarly sharp also upon moving slightly off the $\JP=J$ line. However, in the  transition regime between the DS and the PS phase, the triplet density $n_\text{D}$ exhibits a smooth crossover  in contrast to its sharp jump along the $\JP=J$ line. There are two possible explanations for this observation: (i) 
there exists a finite-temperature first-order transition between both phases, and the line of critical points, along which the wall of discontinuities terminates, resides at temperatures below those accessible to the finite-temperature SSE QMC simulations, or (ii) 
there is no finite-temperature phase transition between the DS and the PS regime, but only a smooth crossover (which however appears unlikely to be realized in a two-dimensional model). In the following section, we will provide arguments from perturbation theory calculations (in $J/\JD$) as well as free-energy considerations that strongly support the first scenario, (i), and derive an explicit expression for $T_c$ along the DS-PS transition line within the perturbative regime. 
 
\section{Perturbation theory and free-energy arguments}\label{Sec:Analytics}

\subsection{Perturbation theory}

Compared to the original FFB, the pFFB model exhibits various weak coupling regimes where perturbation theory can be performed. Here, we are especially interested in the regime where $J\ll \JP\approx \JD$, corresponding to the case of weakly coupled plaquettes. Namely, this regime is contained within the  crossover region observed at finite temperature in QMC (cf. Fig.~\ref{fig:ntrip}). Our goal in the following will be to use perturbation theory in order to understand the physics in this region at low temperatures that are beyond reach of QMC.

\begin{figure}[t]
	\includegraphics{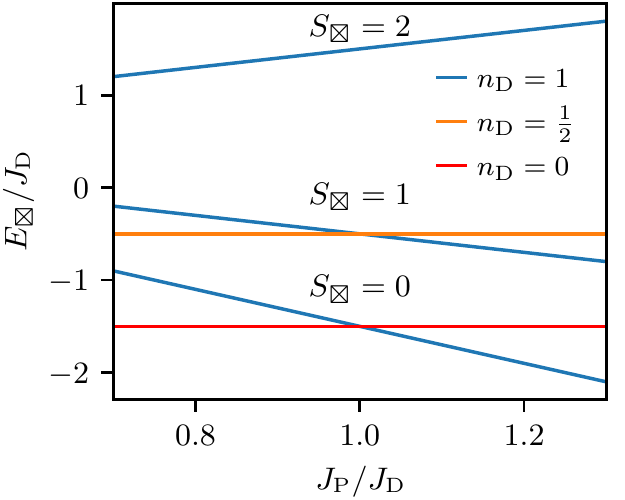}
	\caption{Energy levels of a single plaquette, $E_\boxtimes$, as a function of $\JP/\JD$, splitting into two singlets, three triplets and one quintuplet of different dimer triplet density $\nD$.}
	\label{fig:tetramer}
\end{figure}

To start, we consider the spectrum of a single plaquette, cf.  Fig.~\ref{fig:tetramer}. Based on the symmetries of the problem, the states of the plaquette can be labelled (up to degeneracy) by the plaquette's total spin $\mathbf{S}_{\boxtimes}^2 = (\sum_{i=1}^4 \mathbf{S}_i)^2$ and the dimer triplet density $\nD$. Around $\JP/\JD \approx 1$, the low energy subspace is made up of a dimer singlet state and a plaquette singlet state, which exhibit a level crossing at $\JP/\JD = 1$. From this, we find that in the decoupled limit, i.e.,  for $J=0$, the  pFFB model indeed hosts a level-crossing first-order transition at $T=0$. However, at finite temperature, this transition immediately softens into a crossover at $J=0$.

The next question is what changes in this picture once the interplaquette  interactions $J$ are included. On the level of  quantum numbers, we recall that $\nD$ remains a good quantum number also in the fully coupled model. The same is not true for the other quantum numbers, so the interplaquette interactions will in general mix levels within the different $\nD$-sectors. For the low-energy subspace, this means that the sole $\nD=0$ dimer singlet level, having no other levels to be mixed with, remains the same, while the plaquette singlet level gets shifted, depending on the states on the neighboring plaquettes.

This physics results (see App.~\ref{app:perturb} for a detailed derivation) in an effective low-energy Hamiltonian devoid of any off-diagonal terms,
\begin{align}
	\label{eq:heff_tetra}
	H_\mathrm{eff} &= \sum_\boxtimes \left(\JP - \JD + \frac{5 J^2}{3 \JD}\right) \sigma_\boxtimes^z\nonumber\\
	&- \frac{J^2}{6 \JD} \sigma_\boxtimes^z \left(4 \sigma_{\boxtimes + \hat{x}}^z + \sigma_{\boxtimes+\hat{y}}^z\right) + \mathcal{O}\left(\frac{J^3}{\JD^2}\right),
\end{align}
where we define $\sigma_\boxtimes^z = +1$ ($-1$) if a plaquette is in the dimer (plaquette) singlet state, and $\boxtimes + \hat{x}$ ($\boxtimes + \hat{y}$) denotes the neighboring plaquette to the right (top). This Hamiltonian realizes a classical Ising model with spatially anisotropic interactions in an effective magnetic field. From the expression of the magnetic field, we can read off the existence of a first-order transition along
\begin{equation}
	\label{eq:perturb_foline}
	\JP = \JD - \frac{5 J^2}{3 \JD}+ \mathcal{O}\left(\frac{J^3}{\JD^2}\right), 
\end{equation}
which extends from $T=0$ up to a finite critical temperature $T_c$, which is known from Onsager’s solution \cite{Onsager1944} to satisfy
\begin{equation}
\sinh\left(\frac{2J_x}{T_c}\right) \sinh\left(\frac{2J_y}{T_c}\right) = 1,
\end{equation}
with $J_x=2J^2/3\JD$ and $J_y=J^2/6\JD$ in our case. This yields 
\begin{equation}
T_c \approx 0.826 J^2/\JD
\end{equation}
in the perturbative regime. 
Thus, weak interplaquette couplings are sufficient to stabilize a first-order transition at finite temperature. The line of critical temperature at which the first order transitions terminate is however suppressed by a factor of $J^2$, making it unfeasible to resolve in QMC simulations (e.g., for a value of $J/\JD=0.1$, the above estimates gives $T_c \approx 0.008 \JD$). Nevertheless, we find that the estimate for the first-order transition line  extracted from \eqref{eq:perturb_foline} agrees very well with the position of the $T>T_c$ crossover observed in QMC (cf. the white dashed lines in  Fig.~\ref{fig:ntrip}).

\subsection{Free-energy arguments}

The previous perturbative approach was limited to the region of small $J$, but it was powerful enough to predict the existence and shape of a first-order transition in the thermodynamic limit. In this section, we change our viewpoint, assuming that such a first-order transition exists in the first place and that it happens between two specific quantum number sectors, $\nD=0$ and $\nD=1$. For the weakly coupled regime we just showed that this is the case with the DS ($\nD=0$) and PS phase. For the original FFB this fact is also established with the $\nD=1$ state being an effective $S=1$ AFM~\cite{MullerHartmann2000,Stapmanns2018}. Making this assumption allows us to calculate the first-order line as a level-crossing in the free energy of the two states involved,
\begin{equation}
	F_{\nD=0}(T, \JD, \JP, J) = F_{\nD = 1}(T, \JD, \JP, J),
	\label{eq:freeenergyarg}
\end{equation}
which has been shown to be a very accurate estimate for the shape of the first-order line below the critical temperature in the FFB and related models~\cite{Stapmanns2018,Jimenez2021,Weber2021}. We can further simplify the argument by noting that in the original FFB case the shape of the first-order line depends only weakly on temperature and the same is true for the weakly coupled plaquette regime where the effective Ising magnetic field is independent of temperature. Therefore, we approximate the free energy in Eq.~\eqref{eq:freeenergyarg} by the ground state energy.

The ground state energy of the $\nD=0$ dimer singlet product state,
\begin{equation}
	E_{\nD=0}/N_\text{D} = -\frac{3}{4} \JD,
	\label{eq:End0}
\end{equation}
is known exactly and the equivalent for $\nD=1$ can be written as the sum of the dimer triplet energy and the ground-state energy of the $S=1$ columnar dimer Heisenberg model,
\begin{equation}
	E_{\nD=1}/N_\text{D} = \frac{1}{4} \JD + E_\text{CD}^{S=1}(J, \JP).
	\label{eq:End1}
\end{equation}
The energy $E_\text{CD}^{S=1}(J, \JP)$ is readily accessible to QMC simulations (App.~\ref{app:cdmodel}). It is convenient to introduce an angular parametrization
\begin{align}
	E_\text{CD}^{S=1}(J, J_P) &= E_\text{CD}^{S=1}(J_r \cos \theta, J_r \sin \theta)\nonumber\\&= J_r E_\text{CD}^{S=1}(\theta),
	\label{eq:param}
\end{align}
where due to the structure of the Heisenberg model, the factor $J_r$ can be pulled out. Using these steps, the form of the first-order transition line in the $(J,\JP)$ plane can be written in “polar coordinates” as
\begin{equation}
	J_r = -\frac{\JD}{E_\text{CD}^{S=1}(\theta)}.
	\label{eq:foline}
\end{equation}
The resulting first-order line $J_r(\theta)$ is shown in Fig.~\ref{fig:ntrip} to match the observed crossover and first-order transition across the full phase diagram, in the absence of any fitting parameters. This can be considered indirect evidence for the correctness of our assumptions, and hence for the existence of the first-order transition out of the DS phase for the full range of couplings.

Another point of interest is that Eq.~\eqref{eq:foline} exactly matches the weak-coupling perturbative result from before, as also seen in  Fig.~\ref{fig:ntrip}. This can be understood by remembering that the effect of the perturbations was limited to mixing the different $\nD=1$ levels in the model. These levels all have effective spin $S=1$ and the perturbation theory is thus equivalent to doing perturbation theory for the $S=1$ columnar dimer square lattice model. 

\section{Conclusion}\label{Sec:Conclusions}

\begin{figure}[t]
	\includegraphics{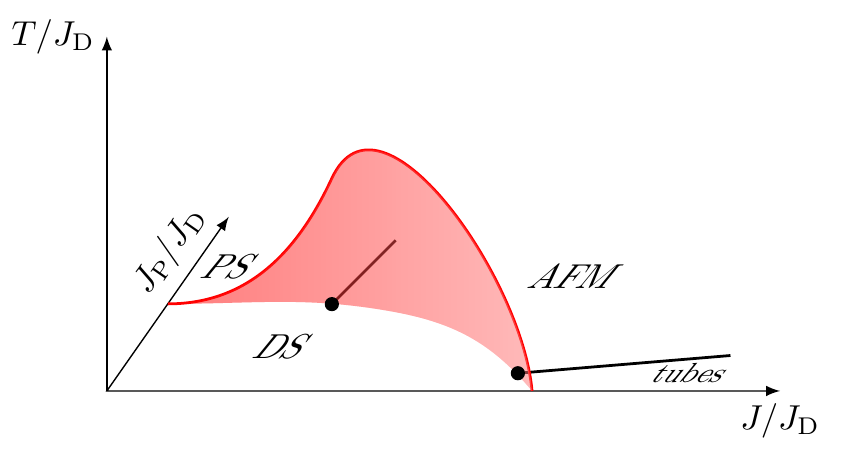}
	\caption{Sketch (not to scale) of the finite temperature phase diagram showing the wall of discontinuities (red) of the finite-temperature first-order transitions in the pFFB model. Near the decoupled plaquette limit, the dependence of the critical temperature $T_c$ (bold red line) on the inter-plaquette coupling is quadratic. Its behavior for small $\JP$ is purely indicative based only on its limiting value $T_c=0$ at $\JP=0$.}
	\label{fig:sketch}
\end{figure}

From a combined analysis using unbiased QMC simulations and perturbation theory as well as free-energy arguments, we  derived the ground state phase diagram of the pFFB spin-1/2 Heisenberg model, and explored in particular the emergence of a line of critical points that terminate a  wall of first-order phase transitions between the DS and the PS low-temperature regimes. A sketch containing both the zero temperature phase diagram as well as the wall of discontinuous first-order transitions and its line of critical points is shown in Fig.~\ref{fig:sketch}.

From the perturbative approach, we derived that the corresponding critical temperature scale in this regime is strongly suppressed by its quadratic dependence on the interplaquette coupling $J$, implying critical scales that fall well below the temperature regime that is accessible to the QMC approach. For the future it might be interesting to provide a similar perturbative approach also in the regime at low $\JP$. Here, a quantum phase transition takes place between the DS and the regime of one-dimensional $J$-tubes, with $T_c = 0$ in the decoupled tube limit $\JP=0$ (how  this limiting value of $T_c$ is approached as $\JP\rightarrow 0$ would be interesting to extract from such a perturbative approach).  Based on the free-energy arguments, we obtained an estimate for the phase boundary of the DS phase that is in remarkable agreement with the results from the QMC simulations. 

We finally note  that the phenomenology observed in the pFFB model, i.e., a discontinous quantum phase transition separating two different quantum-disordered regimes and extending up to a thermal critical point with a comparably low temperature scale, is similar to the thermal physics in SrCu${}_2$(BO${}_3$)${}_2$~\cite{Jimenez2021}. Indeed, the FFB lattice may be considered as a fully frustrated extension~\cite{MullerHartmann2000,Stapmanns2018} of the Shastry-Sutherland model~\cite{Shastry81} that underlies the magnetism in SrCu${}_2$(BO${}_3$)${}_2$. The original FFB model does however not feature a PS phase (in constrast to the Shastry-Sutherland model~\cite{Corboz13}). As we have shown, its plaquettized generalization considered here contains a PS phase and furthermore 
realizes a discontinuous DS-to-PS transition, while its symmetry still protects us from the severe QMC sign-problem that hampers QMC simulations of the Shastry-Sutherland model beyond the DS regime~\cite{Wessel18}. Even though the PS phase of the pFFB model involves no spontaneous symmetry breaking, 
one may nevertheless consider the pFFB model a sign-problem free designer model~\cite{Kaul13}  for the specific thermal physics observed in Ref.~\cite{Jimenez2021} on SrCu${}_2$(BO${}_3$)${}_2$, and considered here.  

\begin{acknowledgments} We thank P. Corboz, A. Honecker and B. Normand for numerous discussions and collaborations on related topics.
We acknowledge support by the Deutsche Forschungsgemeinschaft (DFG) through Grant No. WE/3649/4-2 of the FOR 1807 and through RTG 1995, the Swiss National Science Foundation through Grant No. 182179, the IT Center at RWTH Aachen University and JSC Jülich for access to computing time through the JARA Center for Simulation and Data Science, and the Scientific IT and Application Support Center of EPFL. The Flatiron Institute is a division of the Simons Foundation.
\end{acknowledgments}

\begin{figure}[t]
\includegraphics{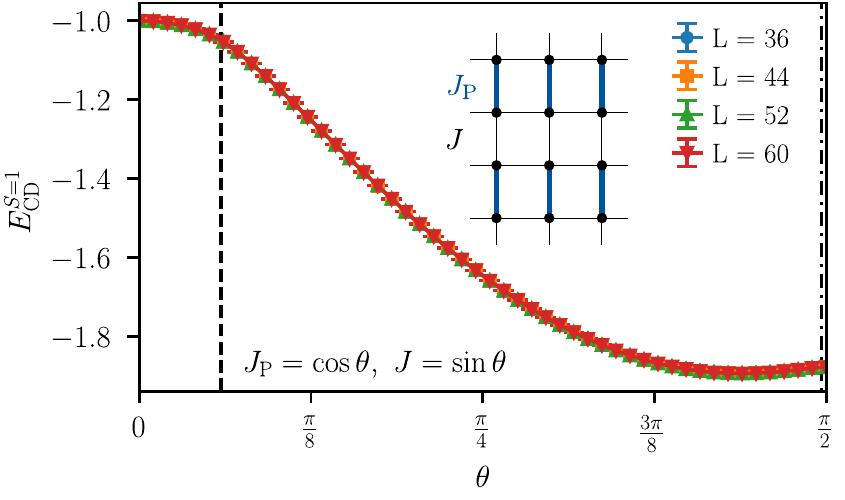}
\caption{Energy per site $E_\text{CD}^{S=1}$ of the $S=1$ Heisenberg model on the columnar dimer lattice (inset) parametrized by the angle $\theta$ for different system sizes $L$. The temperature was scaled as $T=1/2L$ to probe ground state properties.  Black vertical lines denote the boundaries of the central AFM regime from Ref.~\cite{Matsumoto2001}.}
\label{fig:cd_energies}
\end{figure}

\appendix

\section{Details on the perturbative calculation}\label{app:perturb}

In this appendix, we outline the details of the perturbative calculation in the $J \ll \JD \approx \JP$ regime. In this regime the model is well described by weakly coupled plaquettes, and we perform a perturbative downfolding to the low-energy, $S_\boxtimes=0$ sector of each plaquette, consisting of the states
\begin{align}
	\def\u{\uparrow}
	\def\d{\downarrow}
	\ket{\alpha} &= \ket{0,0;0,0},\\
	\ket{\beta} &= \frac{1}{\sqrt{3}} \left(\ket{1,+;1,-}+\ket{1,-;1,+}-\ket{1,0;1,0}\right),
\end{align}
in the dimer basis of the two $\JD$ dimers contained in this plaquette, $\ket{l_1,m_1;l_2,m_2}$.
As discussed in the main text, there are no virtual processes that can renormalize the $\ket{\alpha}$ states so the effective low-energy Hamiltonian can then be written to second order in $J/\JP$ as
\begin{align}
	&H_\text{eff} = \sum_\boxtimes \sum_{p=\alpha,\beta} \varepsilon_p \ketbra{p}{p}_\boxtimes\nonumber\\
	&-\sum_{\braket{\boxtimes,\boxtimes'}} \ketbra{\beta\beta}{\beta\beta}_{\boxtimes,\boxtimes'}\nonumber\\
	&\times\braket{\beta\beta|H_{\boxtimes,\boxtimes'}P\frac{1}{H_\boxtimes+H_{\boxtimes'}-2\varepsilon_\beta}P H_{\boxtimes,\boxtimes'}|\beta\beta},
\end{align}
where
\def\T{\mathbf{T}}
\begin{equation}
	H_\boxtimes = \JP \T_1\cdot \T_2 + \text{const}
\end{equation}
is the single-plaquette Hamiltonian, with
$H_\boxtimes\ket{\alpha} = \varepsilon_{\alpha}\ket{\alpha}$, $H_\boxtimes\ket{\beta} = \varepsilon_{\beta}\ket{\beta}$, and 
\begin{equation}
	P=1-\ketbra{\beta\beta}{\beta\beta} 
\end{equation}
is a projector on the high-energy subspace (all in the $\nD=1$ sector), and lastly
\begin{equation}
	\def\T{\mathbf{T}}
	H_{\boxtimes,\boxtimes'} = J \begin{cases}
		\T_1\cdot \T_{1'} + \T_2\cdot \T_{2'},&\text{if }\boxtimes'=\boxtimes+\hat{x},\\
		\T_1\cdot \T_{2'},&\text{if }\boxtimes'=\boxtimes+\hat{y}.
	\end{cases}
\end{equation}
Here, $\mathbf{T}_{1}$ and $\mathbf{T}_{2}$ ($\mathbf{T}_{1'}$ and $\mathbf{T}_{2'}$) denote the total spins of the two $\JD$ dimers in  plaquette $\boxtimes$ ($\boxtimes'$). The resulting matrix products can be readily evaluated and -- after writing the projectors on $\ket{\alpha}$ and $\ket{\beta}$ in terms of Pauli matrices -- yield the Hamiltonian in Eq.~\eqref{eq:heff_tetra}.

\section{$S=1$ columnar dimer model}\label{app:cdmodel}

In the energy argument for the first-order line, in Eq.~\eqref{eq:foline}, the ground state energy $E^{S=1}_\text{CD}$ of the columnar dimer model (inset of Fig.~\ref{fig:cd_energies}) appears. This quantity is readily accessible in quantum Monte Carlo simulations, which we present in the main panel of Fig.~\ref{fig:cd_energies}.
The QMC simulations were performed using the standard SSE QMC algorithm~\cite{Sandvik1991,Sandvik1999,Syljuasen2002,Alet2016} in the $S=1$ $S^z$-basis. The energy is found to be well converged at $L=60$, where $L$ is the linear system size of the $N=2L^2$ sites spin-1 system.

\section{Higher temperatures}\label{app:higherT}

\begin{figure}[t]
	\includegraphics{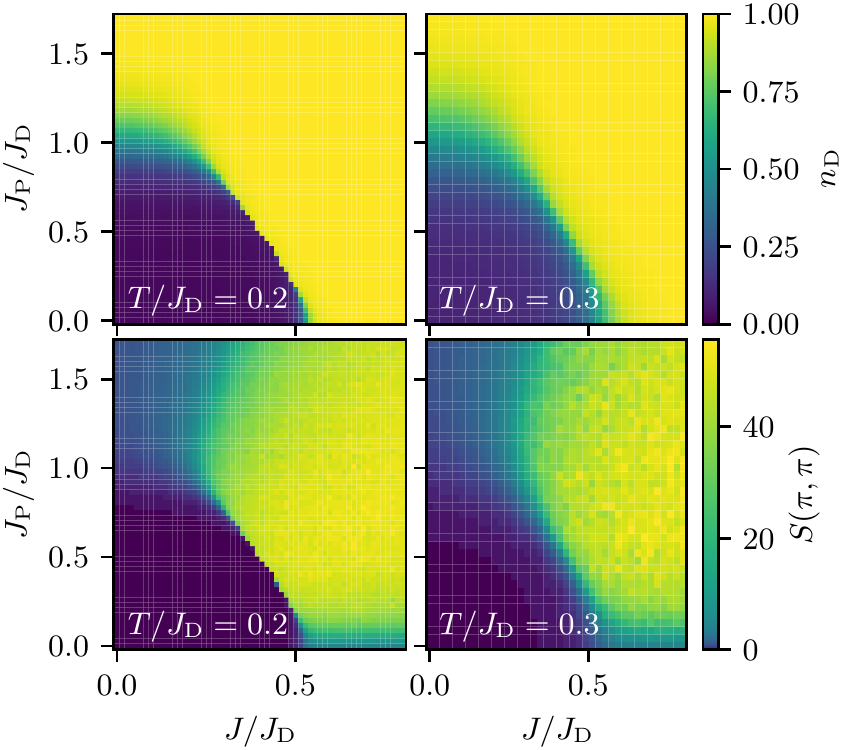}
	\caption{Dimer triplet density $n_\text{D}$ (top panels) and AFM structure factor $S(\pi,\pi)$ (bottom panels) of the pFFB model as functions of $J$ and $\JP$ at fixed temperatures $T/\JD=0.2$ (left panels) and  $T/\JD=0.3$ (right panels) on a system with $L=12$.  }
	\label{fig:higherT}
\end{figure}

In this appendix, we present in Fig.~\ref{fig:higherT} additional results for the dimer triplet denstiy $n_\text{D}$ and the AFM structure factor $S(\pi,\pi)$ taken at higher temperatures. 

\bibliography{paper}
\end{document}